\documentstyle[sprocl]{article}

\input{psfig.sty}

\begin{document}

\title{FORM FACTOR OF THE RELATIVISTIC SCALAR BOUND STATE CALCULATED IN  MINKOWSKI SPACE}

\author{\underline{VLADIM\'{I}R \v{S}AULI}$^{\, a,b}$, J. ADAM, Jr.\ $^a$ }

\address{$^a$ Institute of Nuclear Physics,
  \v{R}e\v{z} near Prague, CZ 250 68, Czech Republic \\
 $^b$ Charles University, Faculty of Mathematics and Physic, Prague,
Czech Republic}

\maketitle\abstracts{We have calculated the electromagnetic elastic form factor
of a two body scalar bound state. The Bethe-Salpeter equation is solved
directly in the Minkowski space using the Perturbation Theory Integral
Representation. At soft coupling regime the obtained results are compared with
those following from a quasipotential spectator (Gross) approximation.}

\section{Introduction}

Relativistic bound states manifest themselves as poles of the scattering matrix
at $P^2= M^2$, where $P$ is the total momentum of the system and $M$ is the
rest mass of the bound state. The residua at those poles define so-called
vertex functions $\Gamma(p,P)$ satisfying the Bethe-Salpeter equation (BSE)
\cite{Nakan1}, which for bound states of two constituents reads:
\begin{equation}
\Gamma(p,P)= i \int\frac{d^4k}{(2\pi)^4}V(p,k,P)G_{0}(k,P)\Gamma(k,P)\, ,
\label{BS}
\end{equation}
where $p$ and $k$ are relative four-momenta of the constituent pair in final
and intermediate states, and the free two-particle propagator $G_{0}(k,P)=
g(k+P/2,m_1) g(k- P/2,m_2) $ is a product of renormalized one body propagators
$g$. The kernel $V$ is a sum of all Bethe-Salpeter irreducible diagrams
\cite{Nakan1}.

For solution, the BSE is usually analytically continued (with the help of the
Wick rotation) into the Euclidean space. This avoids the singularities in the
kernel and propagators and standard numerical techniques can be employed. An
interesting alternative has been developed by Kusaka and Williams
\cite{Kusaka}. It is based on the  Perturbation Theory Integral Representation
(PTIR) \cite{Nakan2}, i.e., on the fact that any n-point Green function can be
expressed as a unique integral over spectral variables. Then, the known
structure of the singularities can be factorized and the BSE is cast into the
real finite integral equation for the real vertex weight function. This
approach is, in principle, applicable also for complicated kernels and/or for
the case when the propagators are arbitrarily dressed, which may make the Wick
rotation difficult or impossible. However, the derivation of the spectral
decomposition is nontrivial and has been so far performed only for 
kernels induced by cubic type - derivative free  interaction of scalars.

In this contribution we present our calculation of the electromagnetic (e.m.)
form factor of a bound state of two scalars, calculated with the help of PTIR.
Our calculations are done with the simplest ladder kernel, only the s-wave
ground state being considered. The results are compared with those obtained in
the 3-dimensional quasipotential (QP) spectator (Gross) reduction of the BSE.

\section{Formalism}

The interaction lagrangian of our simple scalar model is
\begin{eqnarray}
{\cal{L}}_{\rm int}&=& -g\, \psi_1(x)\psi_1(x) \phi(x) - g\, \psi_2^*(x)\psi_2(x)
\phi(x)
 -e^2\, \psi_2^*(x)\psi_2(x) A_{\mu}(x)A^{\mu}(x) \nonumber \\
&& - e\, [ \psi_2^*(x)\partial_{\mu}\psi_2(x)+ \partial_{\mu}\psi_2^* (x)
\psi_2(x) ]A^{\mu}(x)
 \, ,
\end{eqnarray}
where $g$ is a (strong) coupling constant, $ \psi_1$ and $\psi_2$ are fields of
massive scalar particles (with masses $m_1$ and $m_2$), neutral and charged,
respectively, $\phi$ is the lighter scalar field mediating the strong
interaction,  and $ A_{\mu}(x) $ is a photon field. In ladder approximation the
t channel kernel is given by
\begin{equation}
 V(p,k,P)= - \frac{g^2}{ m_{\phi}^2 - (p-k)^2 -i\epsilon } \, .
\label{ladder}
\end{equation}
The spectral decomposition of the bound state vertex function reads
\cite{Kusaka,Nakan2}:
\begin{equation}
\Gamma(p,P)=\int_{0}^{\infty}d\alpha\int_{-1}^{1} \frac{dz\,
\rho^{[2]}(\alpha,z)}{(m^2+ \alpha- (p^2 + z p\cdot P+ \frac{P^2}{4})-
i\epsilon)^2} \, .
\end{equation}
The general scattering kernel can be expressed as a rather complicated
multi-dimensional spectral integral (for details see \cite{Kusaka,Nakan2}).
From the BSE it
then follows for the weight functions:
\begin{equation}
\frac{1}{\lambda}\frac{\rho^{[2]}(\bar{\alpha},\bar{z})}{\bar{\alpha}^{2}}=
\int_{0}^{\infty}d\alpha\int_{-1}^{1}dz \ K^{\rm tot}
(\bar{\alpha},\bar{z},\alpha,z) \ \frac{\rho^{[2]}(\alpha,z)}{\alpha^{2}} \, ,
\label{PTIRE}
\end{equation}
where $ \lambda=g^{2}/(4\pi)^{2}$ and where all the scattering kernel and
propagators weight functions together with appropriate
number of spectral integrals are non-trivially hidden
in $K^{\rm tot}(\bar{\alpha},\bar{z},\alpha,z)$.
This integral equation for $\rho(\alpha ,z)$ has been solved numerically by iterations.

It is interesting to include an annihilation interaction into (\ref{PTIRE}).
Consider a generalized annihilation kernel:
\begin{equation}
V_{\rm ah}(P)= - \int_{0}^{\infty}d\gamma\int_{0}^{1}d\xi \ \frac{\rho_{\rm ah}
(\gamma,\xi)}{\gamma-P^{2}\xi-i\epsilon} \, ,
\end{equation}
which includes a tree level $ V_{\rm ah, tree}= - g^2/(m^{2}_\phi- M^2
-i\epsilon)$. From the assumption that any vertex function can be expressed in
the PTIR form it follows a surprising fact that the annihilation term cannot
contribute to (\ref{PTIRE}), since BSE is derived assuming a zero width of a
bound state. This would mean that the presence of the annihilation term does
not affect the bound state spectrum and wave functions. On the other hand, for
the bound states of $(\psi_1, \psi_1)$ or $(\psi_2, \psi_2)$ the annihilation
channel is open and the bound states decay into two or more $\phi $ particles.
Clearly, in this case the spectrum and the vertex functions are affected by the
annihilation kernel. It  is difficult to take this into account in a self
consistent way, since the usual derivation of the BSE is based on the
factorization of the S-matrix with the real bound state pole, i.e., without the
finite width.

The square of the electromagnetic form factor is an observable quantity. It
appears that the well known expression for the e.m. form factor of the bound
state described by the BSE can be rewritten in terms of the PTIR weight
functions as an integral free of singularities. The resulting expression is
lengthy, but easy to use numerically and will be given explicitly elsewhere.

\begin{figure}[t]

\centerline{ \mbox{\psfig{figure=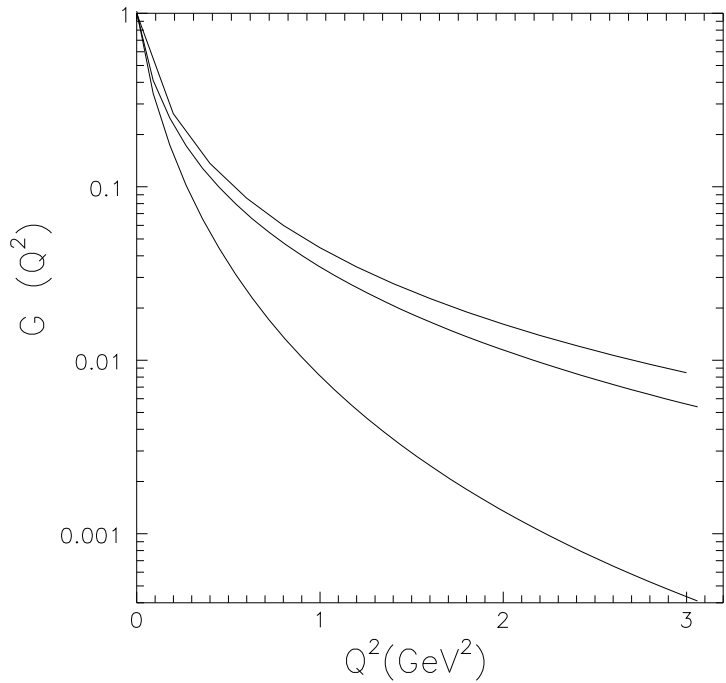,height=9truecm}}              }
\caption[99]{  Form factor of scalar deuteron.
Upper line represents the BS form factor. Two lower lying curves represent
form factors calculated in Gross formalism with (lowest line) and without
(intermediate line) phenomenological strong form factors .
$ Q^2 $ is given in the units of $ m^2 $. }
\end{figure}

\section{Results}
Figure~1a illustrates the differences between Bethe-Salpeter and QP results at
soft coupling regime. We fix the masses of $\psi_1, \psi_2, \phi$ to be equal
to the masses of the neutron, proton and pion, respectively (actually, we use
$m_n=m_p=939\, {\rm MeV},\, m_{\phi}=138\, {\rm MeV} $). The coupling constant
$\lambda$ is for these masses and for $\epsilon_d= 2.3 $ MeV given by $\lambda=
0.118$ (in units of $m_p^2$) . For QP we also introduce usual phenomenological form factors $
{(\Lambda^{2}-m_{\phi}^{2})}/{(\Lambda^{2}-p_{\phi}^{2})} $. Obviously,
Bethe-Salpeter and Gross form factors have very similar shape, the difference
between them is much smaller than the effect due to the modification of the
strong interaction by the phenomenological form factors.

 Figure~1b shows the e.m. form factors for several values of the fraction of binding
$\eta= M/(m_1 + m_2)$ for the choice of masses  $m_1= m_2= m= 2 m_{\phi}$ also
used by Kusaka {\it et al} \cite{Kusaka}. Recall, that for the deuteron $
\eta_d \simeq 0.9988$. Although the fraction of binding varies dramatically,
the form factors look rather similar. It seems that their slopes are much more
sensitive to the range of interaction rather than its strength.

\section{Conclusions and Outlook}

In this contribution the elastic e.m. form factors of scalar bound states have
been calculated in Bethe-Salpeter and Gross formalism. We also claim that
annihilation term does not influence the spectrum and wave functions of the
bound states.

 For more realistic systems the complication of spin degrees of
freedom has to be considered. We have formally derived similar equation for
fermion-antifermion state. Its solution is being developed.

\begin{figure}[t]
\centerline{  \mbox{\psfig{figure=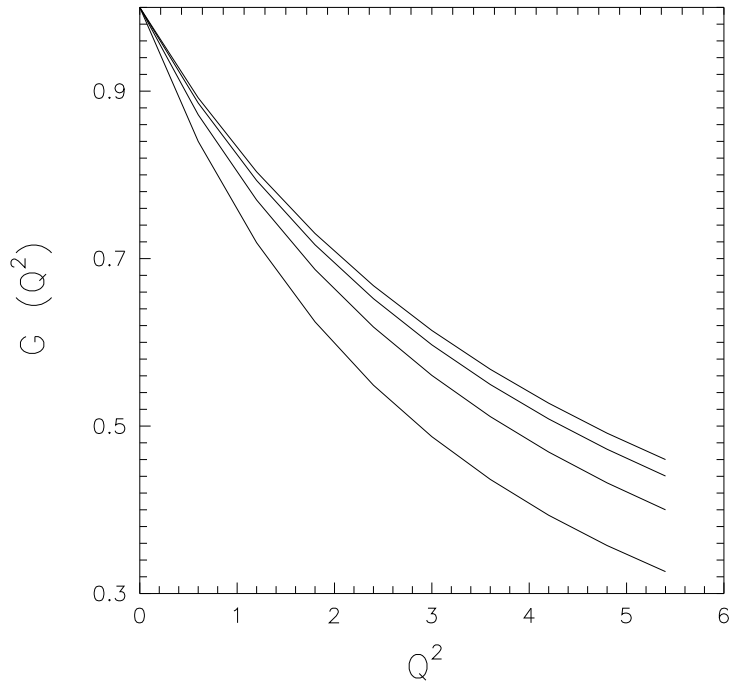,height=9truecm}} }
\caption[99]{ The BS e.m. form factors for the following fraction of binding: $
\eta= 0.1, 0.2, 0.6, 0.8$.  $ Q^2 $ is given in the units of
$ m^2 $. }
\end{figure}

\section*{Acknowledgements}
This work was supported by the grant GA CR 2020447. Some of the results were
part of the Diploma thesis of V.~\v{S}.

* added in November 2001


\section*{References}

\begin{thebibliography}{99}

\bibitem{Nakan1}
 see. e.g., N.~Nakanishi, {\it Prog. Theor. Phys., Suppl.} {\bf 43}, 1 (1969)
  and references therein

\bibitem{Kusaka}
 K.~Kusaka, K.~Simpson, A.G.~Williams, {\it Phys. Rev.} D {\bf 56}, 5071 (1997)
and references therein, The authors of this work used
$\ K(\bar{\alpha},\bar{z},\alpha,z)$ which was derived as at least one dimensional 
certain integral representation. It was shown by one of the author (V.\v{S}.)
that this integration can be performed analytically. Avoiding this additional integration 
significantly decrease the necessary computer time. The appropriate statement will be
prooved in the article which one is under preparation.* 

\bibitem{Nakan2}
 N.~Nakanishi, {\it Graph Theory and Feynman Integrals.}, (Gordon and
Breach, New York, 1971).

\bibitem{Wally}
  F.~Gross, {\it Phys. Rev.} {\bf 186}, 1448 (1969);
  {\it Phys. Rev.} D {\bf 10}, 223 (1974); {\it Phys. Rev.} C {\bf 26}, 2203 (1982); \\
  J.W.~Van~Orden, {\it Czech. Jour. of Phys.} B {\bf 45}, 181 (1995)

%\bibitem{Sauli}
%  Vladimir \v{S}auli, in preparation.

\end{thebibliography}
\end{document}